\documentclass{INTERSPEECH2023}
\usepackage{color,amsmath,url,times,tabularx,bbm,amssymb,multirow}
\usepackage{bm}
\usepackage{cite}

\interspeechcameraready

\title{Masked Audio Modeling with CLAP and Multi-Objective Learning}
\name{Yifei Xin$^{1 \ast}$,
      Xiulian Peng$^{2}$,
      Yan Lu$^{2}$\thanks{$^{\ast}$This work was done at Microsoft Research Asia.}
      }
\address{
  $^{1}$Peking University, Beijing, China\\
  $^{2}$Microsoft Research Asia, Beijing, China}
\email{xinyifei@stu.pku.edu.cn}

\begin{document}

\maketitle
\begin{abstract}
Most existing masked audio modeling (MAM) methods learn audio representations by masking and reconstructing local spectrogram patches. However, the reconstruction loss mainly accounts for the signal-level quality of the reconstructed spectrogram and is still limited in extracting high-level audio semantics. In this paper, we propose to enhance the semantic modeling of MAM by distilling cross-modality knowledge from contrastive language-audio pretraining (CLAP) representations for both masked and unmasked regions (MAM-CLAP) and leveraging a multi-objective learning strategy with a supervised classification branch (SupMAM), thereby providing more semantic knowledge for MAM and enabling it to effectively learn global features from labels. Experiments show that our methods significantly improve the performance on multiple downstream tasks. Furthermore, by combining our MAM-CLAP with SupMAM, we can achieve new state-of-the-art results on various audio and speech classification tasks, exceeding previous self-supervised learning and supervised pretraining methods.
\end{abstract}
\section{Introduction}
\label{sec:intro}
Recent years have seen considerable success in masked image modeling (MIM) \cite{he2022masked,xie2022simmim,zhang2022cae} in the image domain, demonstrating promising results on a variety of downstream tasks, such as image classification, semantic segmentation, and object detection. MIM uses a pre-defined mask ratio to mask out image patches and adds the reconstruction supervision to the masked regions. In this study, we investigate the learning of general audio representations using MIM applied to the audio spectrogram, which we call masked audio modeling (MAM) \cite{huangmasked,baade2022mae,chong2022masked}. MAM splits the audio spectrogram into patches along the time and frequency axes \cite{koutini2021efficient,chen2022hts,xin2023enhancement}, enabling the model to learn both temporal and frequency structures.

Currently, MAM has been applied to general audio represenation learning, showing good performance on audio and speech understanding tasks \cite{gong2022ssast,xin2022low,xin2023background,xin2023causal}. This indicates that MAM learns robust auditory representations for both speech and non-speech signals. However, it is generally believed that the reconstruction loss mainly accounts for signal-level quality at time-frequency domain but is still limited in learning high-level audio semantic knowledge \cite{baobeit}. Besides, audio exhibits different characteristics from images. It is not easy to find an auditory-perception-aware patching strategy that fits various sound events with diverse scales. Masking with an extremely high ratio (80\% mask ratio on AudioMAE \cite{huangmasked}) would unavoidably harm sound events with a short time/frequency span. 

In addition to MAM, supervised pretraining is also widely studied for audio semantic modeling, which leverages either out-of domain data (e.g. ImageNet \cite{deng2009imagenet}) or in-domain audio data (e.g. AudioSet \cite{gemmeke2017audio}) for pretraining. As for the out-of-domain supervised pretraining, HTS-AT \cite{chen2022hts} employs the Swin Transformer-based architecture \cite{liu2021swin} as the backbone, which obtains superior audio understanding performance \cite{wu2023large,xin2023cooperative}. For in-domain supervised pretraining, inspired by the language-vision pretraining method CLIP \cite{radford2021learning}, CLAP \cite{elizalde2022clap} utilizes a contrastive language-audio pretraining task \cite{mei2023wavcaps,xin2023retrieval,xin2023interaction} to learn text-enhanced audio representations with audio and text pairs. Compared with MAM, these methods learn different levels of semantics, which could complement each other. 

In this paper, we aim to enhance the semantics for MAM by leveraging external cross-modal information. To this end, we present MAM-CLAP, a simple yet effective framework that incorporates cross-modal information into MAM. We take AudioMAE \cite{huangmasked} as the benchmark for MAM, which learns to efficiently encode the visible patches into latent representations that carry essential information for reconstructing masked patches. Our motivation is that models trained with cross-modal data can provide additional richer semantic knowledge. In this case, we apply the CLAP model \cite{elizalde2022clap} to provide semantic guidance for MAM as it shows great robustness on various audio downstream tasks. Moreover, unlike most MAM methods applying the reconstruction supervision on masked patches, we find that supervisions on both masked and visible patches with the CLAP target can achieve remarkable performance. It demonstrates that the visible patches can effectively extract rich semantic information from CLAP, performing like the feature distillation.
\begin{figure*}[t]
  \centering
  \includegraphics[width=1.0\linewidth]{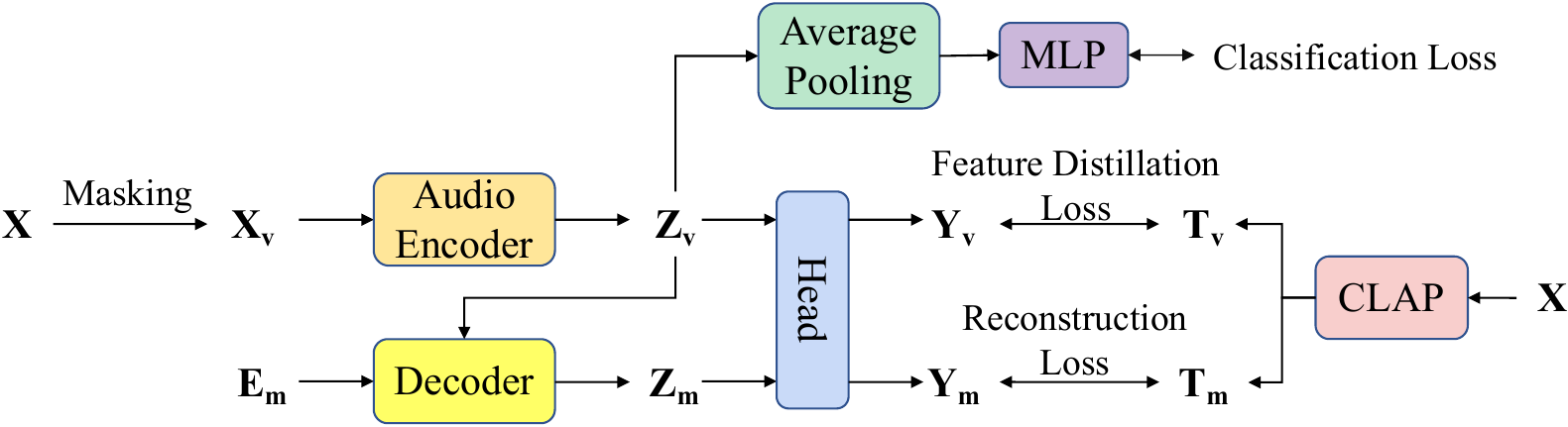}
  \caption{The overview of our SupMAM-CLAP.}
  \label{fig:adapt-pic}
  \vspace*{-\baselineskip}
\end{figure*}

Furthermore, most existing MAM methods \cite{huangmasked,baade2022mae,niizumi2022masked} only learn feature interactions among patches. No global features, i.e., features that can represent the entire spectrogram, are explicitly learned during pretraining. To complement this, we extend MAM to a supervised setting (SupMAM) by adding a branch for supervised classification in parallel with the reconstruction target, thereby enabling MAM to effectively learn global features from labels. To make it sample-efficient, only a subset of visible patch features are used for the classification-based pretraining branch, where an average pooling is employed to get the global audio representation followed by a MLP for classification. 

In a nutshell, our contributions are threefold: 
\begin{itemize}
\item We provide a novel perspective that the CLAP-targeted feature supervision on the spectrogram patches is a good choice for audio representation learning. We develop a simple yet effective framework MAM-CLAP to improve the semantic modeling of MAM-based audio pre-training.  
\item We introduce a multi-objective learning method that extends AudioMAE to a fully-supervised setting (SupMAM) by adding a supervised classification branch. SupMAM only uses a subset of the visible patches to do classification rather than all patches used in standard supervised pre-training.
\item By combining our MAM-CLAP with SupMAM, our SupMAM-CLAP sets new state-of-the-art performance across various audio and speech classification benchmarks.
\end{itemize}

\begin{table*}
  \caption{Comparing with the SOTA models on audio and speech classification tasks. IN, AS, and LS denote the ImageNet, AudioSet, and LibriSpeech datasets, respectively. TA denotes the text-audio pairs for CLAP pretraining.}
  \vspace*{-\baselineskip}
  \centering
  \label{tab:freq}
  \begin{tabular}{lcccccl}
    \toprule
    Method & Data & AS-2M & AS-20K & ESC-50 & KS1 & KS2\\
    \midrule
    \textbf{Out-of-domain Supervised Pretraining} & & & & & &\\
    PSLA \cite{gong2021psla} & IN & 44.0 & 31.9 & - & - & 96.3\\
    AST \cite{gong2021ast} & IN & 45.9 & 34.7 & 88.7 & 95.5 & 98.1\\
    PaSST \cite{koutini2021efficient} & IN+AS & 47.1 & - & - & - & -\\  
    HTS-AT \cite{chen2022hts} & IN & 47.1 & - & - & - & 98.0\\
    \midrule
    \textbf{In-domain Supervised Pretraining} & & & & & &\\
    PANN \cite{kong2020panns} & AS & - & - & 94.7 & - & -\\
    PaSST \cite{koutini2021efficient} & IN+AS & - & - & 96.8 & - & -\\
    HTS-AT \cite{chen2022hts} & IN & - & - & 97.0 & - & -\\
    CLAP \cite{elizalde2022clap} & TA & - & - & 96.7 & - & -\\   
    \midrule
    \textbf{Self-Supervised Pre-Training} & & & & & &\\
    SS-AST \cite{gong2022ssast} & AS+LS & - & 31.0 & 88.8 & 96.0 & 98.0\\
    MSM-MAE \cite{niizumi2022masked} & AS & - & - & 85.6 & - & 87.3\\
    MaskSpec \cite{chong2022masked} & AS & 47.1 & 32.3 & 89.6 & - & 97.7\\
    MAE-AST \cite{baade2022mae} & AS+LS & - & 30.6 & 90.0 & 95.8 & 97.9\\ 
    AudioMAE \cite{huangmasked} & AS & 47.3 & 37.1 & 94.1 & 96.9 & 98.3\\
    \midrule
    \textbf{Ours} & & & & & &\\
    MAM-CLAP & AS+TA & \textbf{48.0} & \textbf{38.2} & \textbf{97.1} & \textbf{97.7} & \textbf{98.5}\\
    SupMAM & AS & \textbf{47.7} & \textbf{37.6} & \textbf{96.2} & \textbf{97.3} & \textbf{98.4}\\
    SupMAM-CLAP & AS+TA & \textbf{48.5} & \textbf{38.6} & \textbf{97.6} & \textbf{98.0} & \textbf{98.7}\\
  \bottomrule
\end{tabular}
\vspace*{-\baselineskip}
\end{table*}
\section{SupMAM-CLAP}
\label{sec:proposed method}
\subsection{MAM-CLAP}
The overview of our SupMAM-CLAP is illustrated in Figure 1. Let $X$ denote an input spectrogram. We take AudioMAE \cite{huangmasked} as the backbone. Our MAM-CLAP first embeds $X$ into $N$ patches, which are then randomly masked by a specific proportion $\gamma$. These $N$ patches are naturally split into two non-overlapped sets, i.e., visible patches $X_v$ and masked patches $X_m$, where $N = |v| + |m|$. The mask ratio is thus denoted as $\gamma = |m| / N$. The AudioMAE encoder $e(\cdot)$ maps the visible patches $X_v$ to the latent representations $Z_v$. The decoder $g(\cdot)$ predicts the latent representations $Z_m$ for the masked patches from mask tokens $E_m$, conditioned on the visible latent representation $Z_v$. After that, the predictions of visible patches $Y_v$ and masked patches $Y_m$ are obtained via a head $h(\cdot)$. In this work, we only use a fully-connected layer followed by a layer normalization in $h(\cdot)$. For the target supervision, we directly input the spectrogram $X$ into the CLAP model $f(\cdot)$ to generate the target supervision $T$. $T$ is then split into $T_v$ and $T_m$ corresponding to the positions of $X_v$ and $X_m$. The optimization is applied on $Y_v$ to approach $T_v$, and we also add the supervision on $Y_m$ using $T_m$.

\subsection{SupMAM}
Unlike standard supervised pretraining methods \cite{gong2021ast,kong2020panns,xin2022audio} that use all patch features, SupMAM only uses a subset of the visible patches to do classification, which makes SupMAM more sample-efficient. This is mainly based on the intuition that humans can recognize sound events with partial information (a subset of patches). Also, from the perspective of data augmentation, random masking can generate different training samples for each iteration \cite{baade2022mae,liang2022supmae}, serving as a strong regularization.

In this case, a global pooling first condenses the visible local patch features $Z_v$ into the global representation $G_v$ of the spectrogram, which is then used to predict the sound event labels. The classification branch is complementary to the feature reconstruction branch as it can bring global feature learning into the framework. For the classification head, we use a two-layer MLP, with a batch normalization and a ReLU activation injected in-between to project the global representation into the logits of the predicted probabilities $p_v$. During fine-tuning, we only use the AudioMAE-based Transformer encoder for downstream tasks. 

\subsection{Loss Functions}
Most previous MAM methods \cite{huangmasked,chong2022masked,niizumi2022masked} only apply the reconstruction supervision on the predictions of masked patches. With CLAP as the target, we supervise both visible and masked patches. The L2 loss functions are used as follows:
\begin{equation}
\begin{aligned}
    \mathcal{L}_{target} = \frac{1}{l_v}{\Vert Y_v, T_v\Vert}_2 + \frac{1}{l_m}{\Vert Y_m, T_m\Vert}_2,
\end{aligned}
\end{equation}
where $l_v$ and $l_m$ denote the number of visible and masked tokens.

For SupMAM, we use the cross-entropy (CE) loss $\mathcal{L}_{cls}$ to supervise the training process with the classification label $y$:
\begin{equation}
\begin{aligned}
    \mathcal{L}_{cls} = CE(p_v/\tau, y),
\end{aligned}
\end{equation}
where $\tau$ is a temperature parameter that controls the concentration level of the distribution, which is widely used in supervised \cite{wang2017normface} and self-supervised \cite{wu2018unsupervised} feature learning. $\tau$ is set to 10 in our experiments.

Finally, our SupMAM-CLAP is optimized with both the CLAP-target loss and the classification loss, which simultaneously learns fine-grained local and global features. We use a weighted sum of these two loss terms as our overall loss as follows:
\begin{equation}
\begin{aligned}
    \mathcal{L}_{total} = \mathcal{L}_{target} + \lambda_{cls} \mathcal{L}_{cls},
\end{aligned}
\end{equation}
where $\lambda_{cls} $ are weights to balance the objective. Note that when only using our SupMAM framework, $\mathcal{L}_{target}$ is replaced by the mean square error (MSE), i.e., the reconstruction loss of AudioMAE on the masked portion of the reconstructed and the input spectrogram. 
\begin{table}
  \caption{Influences of the mask ratio in our MAM-CLAP.}
  \vspace*{-\baselineskip}
  \centering
  \label{tab:freq}
  \begin{tabular}{ccl}
    \toprule
    Mask Ratio & AS-20K & ESC-50\\
    \midrule
    10\% & 38.1 & 96.9\\
    \textbf{20\%} & \textbf{38.2} & \textbf{97.1}\\
    30\% & 37.9 & 96.9\\
  \bottomrule
\end{tabular}
\vspace*{-\baselineskip}
\vspace*{-0.1cm}
\end{table}

\section{Experiments}
\label{sec:exp}
We perform extensive evaluations on five tasks, including audio classification on AudioSet (AS-2M, AS-20K) \cite{gemmeke2017audio} and Environmental Sound Classification (ESC-50) \cite{piczak2015esc}, and speech classification on Speech Commands (SPC-1 and SPC-2) \cite{warden2018speech}. 

\subsection{Datasets}
\textbf{AudioSet (AS-2M, AS-20K)} \cite{gemmeke2017audio} contains about 2 million 10-second YouTube clips with 527 sound events. The full training set has 2 subsets: a class-wise balanced (22,176 clips) and an unbalanced (2,042,985 clips) set. The evaluation set has 20,383 clips. We collected and processed around 1.9M unbalanced training, 21K balanced training, and 19K evaluation clips due to the frequent change in YouTube videos available (e.g., videos being removed or taken down). For the AS-2M experiments, we use all unbalanced and balanced training audio clips for pretraining and finetuning. For the AS-20K experiments, we employ AS-2M for pretraining and the 20K balanced set for fine-tuning. We evaluate our models on the 19K evaluation set with the mean average precision (mAP) evaluation metric.

\textbf{Environmental Sound Classification (ESC-50)} \cite{piczak2015esc} is an audio classification dataset including 2,000 5-second environmental sound recordings annotated with 50 classes. Each recording is only tagged with one class. We follow the 5-fold cross-validation evaluation setting as the previous work \cite{huangmasked} and report the classification accuracy as the evaluation metric.

\textbf{Speech Commands (SPC-2, SPC-1)} \cite{warden2018speech} are two keyword spotting tasks. There are 35 speech commands in SPC-2. The training/validation/testing set contain 84,843/9,981/11,005 1-second recordings, respectively. In SPC-1, there are 10 classes of keywords, 1 silence class, and 1 unknown class that includes all the other 20 common speech commands. We report the testing accuracy based on the data and split provided by the SUPERB benchmark \cite{yang21c_interspeech}.

\begin{table}
  \caption{Influences of the decoder block in our MAM-CLAP.}
  \vspace*{-\baselineskip}
  \centering
  \label{tab:freq}
  \begin{tabular}{ccl}
    \toprule
    Decoder Block & AS-20K & ESC-50\\
    \midrule
    \textbf{1} & \textbf{38.2} & \textbf{97.1}\\
    2 & 37.8 & 96.8\\
    3 & 37.6 & 96.5\\
  \bottomrule
\end{tabular}
\vspace*{-\baselineskip}
\vspace*{-0.1cm}
\end{table}
\subsection{Training Details}
We employ a standard 12-layer ViT-B by default as the Transformer encoder, where we keep the model size similar to AudioMAE for a fair comparison. For the decoder, we only use a single Transformer block with shifted local attention. In addition, we follow the training pipeline of AudioMAE \cite{huangmasked} to train our models. We resample each raw waveform to 16 khz, and extract the 128-dimensional mel-filter bank features using a 25ms Hanning window that shifts every 10ms as the acoustic feature. The resulting spectrogram for a 10-second clip in AudioSet has a dimension of 1 × 1024 × 128. Each acoustic feature is divided into 16 × 16 patches, which are then flattened into a patch sequence as the model input. We use AudioSet-2M for pretraining and randomly iterate over all audio clips. We distribute the training load over 4 V100 GPUs with a batch size of 128 and a learning rate of 0.0002. 

For our MAM-CLAP, we adopt the CLAP feature to supervise the masked and unmasked patches. For our SupMAM, we leverage the patch reconstruction branch and the supervised classification branch simultaneously. For our SupMAM-CLAP, we use the CLAP feature to supervise the masked and unmasked patches while employing the supervised classification branch. $\lambda_{cls}$ is set as 0.01 for SupMAM and 0.0001 for SupMAM-CLAP. Different from AudioMAE with a 80\% mask ratio during pretraining, we use a masking ratio of 20\% with unstructured random masking for our MAM-CLAP, and adopt a 40\% masking ratio for our SupMAM and SupMAM-CLAP to achieve better performance. During pretraining, we do not apply any augmentation methods. During the finetune stage, we employ a structured random masking ratio (0.2 in time and 0.2 in frequency) for our methods.

\subsection{Experimental Results}
Table 1 compares our methods to prior state-of-the-art. With the CLAP target, our MAM-CLAP achieves the best performance across all tasks compared to previous models with self-supervised and supervised pretraining methods. On AudioSet-2M, it improves the performance to 48.0 mAP. On AudioSet-20K and ESC50, MAM-CLAP outperforms AudioMAE by a large margin. For the speech tasks (KS1, KS2), MAM-CLAP also achieves competitive performance with prior SOTA methods. What’s more, our SupMAM also demonstrates superior performance on multiple downstream tasks. By combining our MAM-CLAP with SupMAM, we can achieve new SOTA results on all tasks compared to all previous SOTA models, which strongly demonstrates the effectiveness of our methods.

\begin{table}
  \caption{Influences of the mask ratio in our SupMAM.}
  \centering
  \vspace*{-\baselineskip}
  \label{tab:freq}
  \begin{tabular}{ccl}
    \toprule
    Mask Ratio & AS-20K & ESC-50\\
    \midrule
    30\% & 37.5 & 96.1\\
    \textbf{40\%} & \textbf{37.6} & \textbf{96.2}\\
    50\% & 37.5 & 96.0\\
  \bottomrule
\end{tabular}
\end{table}

\begin{table}
  \caption{Influences of pretraining objectives in our SupMAM.}
  \centering
  \vspace*{-\baselineskip}
  \label{tab:freq}
  \begin{tabular}{ccl}
    \toprule
    Pretraining Objectives & AS-20K & ESC-50\\
    \midrule
    rec & 36.6 & 93.3\\
    cls & 33.9 & 91.9\\
    rec+cls & \textbf{37.6} & \textbf{96.2}\\
  \bottomrule
\end{tabular}
\vspace*{-\baselineskip}
\end{table}

\subsection{Ablation Study}
In this part, we discuss the influence of the mask ratio and the decoder block in our MAM-CLAP, the mask ratio, the pretraining targets, the classification loss ratio in our SupMAM, and the classification loss ratio, the mask ratio in our SupMAM-CLAP. Here, we test the performance on the AS-20K and ESC-50 datasets.

\textbf{Influences of the mask ratio in our MAM-CLAP.} 
Given that we replace the reconstruction target with the CLAP feature and apply the supervision on both masked and visible patches in MAM-CLAP, we suppose that it may not be appropriate to adopt a high mask ratio as that in AudioMAE. Here, we show the results of different mask ratios. It can be seen that unlike AudioMAE that achieves the best score with 80\% mask ratio, our MAM-CLAP with 20\% mask ratio achieves the best performance.

\textbf{Influences of the decoder depth in our MAM-CLAP.} As show in Table 3, since we do not need to use the decoder for the patch reconstruction task to process the signals of specified scale, a single block decoder can achieve better accuracy than the heavier one with multiple blocks, which is significantly more efficient compared with AudioMAE with 16 decoder blocks.

\textbf{Influences of the mask ratio in our SupMAM.} In this part, we extend AudioMAE to a supervised setting. As shown in Table 4, too large mask ratio will decrease the performance under supervised setting. The reason may be that when there are too many patches masked out, some short or inconspicuous sound events are easily erased up, resulting in incorrect classification and adversely affecting the model training. A 40\% mask ratio can bring the best performance while retaining the advantages of AudioMAE and the supervised classification branch.

\textbf{Influences of the pretraining objectives in our SupMAM.} Table 5 studies the pretraining objectives with 40\% mask ratios. The method degrades into AudioMAE with only the reconstruction (rec) objective. If only the classification (cls) objective is used, the method degrades into standard supervised pretraining with 40\% input patches masked out. We discover that neither the reconstruction nor the classification objective can perform well when used in isolation. This is because only with both objectives can 100\% patches be exploited: (1) reconstruction operates on the masked patches, and (2) classification operates on the visible patches.

\textbf{Influences of the classification loss ratio in our SupMAM and SupMAM-CLAP.} As shown in Table 6, to achieve a balance between the two pretraining goals for SupMAM, we first set the ratio of the reconstruction loss to 1, and then tune the classification ratio. We find that a 0.01 classification loss ratio is optimal. A ratio that is too high would cause SupMAM to degrade into standard supervised pretraining (with a large proportion of input patches masked out), which is detrimental to the network pretraining. For SupMAM-CLAP, we set the ratio of our target loss to 1. It can be seen that a 0.0001 classification loss ratio performs best.

\begin{table}
  \caption{Effects of the classification loss ratio in our SupMAM and SupMAM-CLAP.}
  \centering
  \vspace*{-\baselineskip}
  \label{tab:freq}
  \begin{tabular}{ccl}
    \toprule
    Classification Loss Ratio & AS-20K & ESC-50\\
    \midrule
    SupMAM & &\\
    0.02 & 37.5 & 96.0\\
    \textbf{0.01} & \textbf{37.6} & \textbf{96.2}\\
    0.005 & 37.3 & 95.8\\
    \midrule
    SupMAM-CLAP & &\\
    2e-4 & 38.2 & 97.3\\
    \textbf{1e-4} & \textbf{38.6} & \textbf{97.6}\\
    5e-5 & 38.5 & 97.4\\
  \bottomrule
\end{tabular}
\end{table}

\begin{table}
  \caption{Influences of the mask ratio in our SupMAM-CLAP.}
  \centering
  \vspace*{-\baselineskip}
  \label{tab:freq}
  \begin{tabular}{ccl}
    \toprule
    Mask Ratio & AS-20K & ESC-50\\
    \midrule
    30\% & 38.4 & 97.3\\
    \textbf{40\%} & \textbf{38.6} & \textbf{97.6}\\
    50\% & 38.2 & 97.2\\
  \bottomrule
\end{tabular}
\vspace*{-\baselineskip}
\end{table}

\textbf{Influences of the mask ratio in our SupMAM-CLAP.} Table 7 presents the results of our SupMAM-CLAP with different mask ratios. It can be seen that a 40\% mask ratio can bring the best performance, which combines the advantages of both cross-modality semantic guidance and classification supervision.

\section{Conclusions}
\label{sec:conclusion}
In this paper, we propose to enhance the semantic modeling of MAM by distilling cross-modality knowledge from CLAP representations for both masked and unmasked regions (MAM-CLAP) and leveraging a multi-objective learning strategy with a supervised classification branch (SupMAM), thereby providing more semantic knowledge for MAM and enabling it to effectively learn global features from labels. Experiments show that our MAM-CLAP significantly improves the performance on multiple downstream tasks. Moreover, by combining our MAM-CLAP with SupMAM, we can achieve state-of-the-art results on various audio and speech classification benchmarks.
\bibliographystyle{IEEEtran}
\bibliography{mybib}

\end{document}